\documentclass[prl,twocolumn,floatfix,showpacs,superscriptaddress]{revtex4}
\usepackage{amsmath}
\usepackage{graphicx}

\begin{document}

\title{Impurity effects in the quantum Kagome system ZnCu$_{3}$(OH)$_6$Cl$_2$ ?} 
\author{R. Chitra}
\affiliation{Laboratoire de Physique Theorique de la Mati\`ere 
Condense\'e, UMR 7600, Universite de Pierre et Marie Curie, Jussieu, Paris-75005, France.}
\author{M. J. Rozenberg}
\affiliation{Laboratoire de Physique des Solides, CNRS-UMR8502, Universit\'e de Paris-Sud,
Orsay 91405, France.}
\affiliation{Departamento de F\'{\i}sica, FCEN, Universidad de Buenos Aires,
Ciudad Universitaria Pab.I, (1428) Buenos Aires, Argentina.}
 
\date{today}
\begin{abstract}
Motivated by the recent experiments on 
the new spin half Kagome compound
 ZnCu$_{3}$(OH)$_6$Cl$_2$, we study a
phenomenological model of a frustrated quantum magnet.
The model has a spin liquid groundstate and
is constructed so as to mimic the macroscopically
large quasi-degeneracies expected in
the low-lying energy structure of a Kagome system.
We use numerical studies of finite size systems
to investigate the  static as well
as the dynamical response at finite temperatures.
The results obtained using our simple model  are compatible with 
a large number of recent experiments 
including neutron scattering data.
Our study suggests  that 
many of the anomalous  features observed in  experiments have
a natural interpretation in terms of a  spin-$\frac12$
defects (impurities) coupled to an underlying Kagome-type spin liquid.

\end{abstract}


\maketitle

The quest for spin liquids  in real materials  has been in the forefront of
condensed matter physics since Anderson's suggestion of resonating
valence bond   states as a possible  ground state  for the antiferromagnetic
insulating phase seen in high Tc superconductors.
Typically one expects spin liquid ground states in  geometrically frustrated
systems  with a high degree of frustration like in the Heisenberg antiferromagnet
on pyrochlore lattices, checkerboard lattices and the Kagome 
lattice \cite{lhuillier-review}.
The  Heisenberg antiferromagnet on the Kagome lattice  is particularly interesting in 
that numerical calculations show the existence of a special kind of spin liquid ground
state with a macroscopic quasi degeneracy\cite{Kagome1}. 
However, not much is known about the dynamical properties of these spin liquids.
Interest in the physics of Kagome lattices has been boosted by the
synthesis of a novel paratacamite compound ZnCu$_{3}$(OH)$_6$Cl$_2$, which is considered 
as a faithful realization of a  spin-$\frac 12$ Kagome system \cite{physicstoday}.
A spate of recent experiments \cite{helton,ofer,mendels,imai} have however, 
generated a huge debate about whether the observations
are actually due to intrinsic Kagome physics, or whether other aspects like
 impurities or Dzyaloshinski-Moriya interactions\cite{rrpsingh}  may
need to be taken into account. 

Motivated by the current lack of perfect stoichiometric control in the synthesis
of ZnCu$_{3}$(OH)$_6$Cl$_2$, and its impurity-like low temperature 
behavior of the static magnetic susceptibility, we introduce a schematic
model which toys with the idea that the anomalous features seen in experiments 
may arise from spinful impurities dressed by an underlying spin liquid with
an energy spectrum of the Kagome type. Our  mean-field model  explicitly mimics the  principal feature of the Kagome antiferromagnet: a  high degree of
frustration leading to a  macroscopic degeneracy of the ground state. Due to its
mean-field nature, our model is much easier to study than the full Kagome system.
We  begin by considering the following hamiltonian,
\begin{equation}
 H= \sum_{i\neq j=1}^N J_{ij} {\bf S}_i \cdot {\bf S}_j
\end{equation}
where, ${\bf S}$ represent the spin half operators and $J_{ij}$ the antiferromagnetic
interactions between two spins.  
Since all the spins interact with each other, the model is
maximally frustrated. For $J_{ij}=J$ for all $i,j$, the model is exactly soluble
since the hamiltonian can be written as 
$H= J/2[ {\bf S}_{tot}^2 - 3N/4]$ and all states with a given
value of total spin ${ S}_{tot}$ are degenerate. 
For $N$ even, the ground state is  a bunch of
degenerate singlets with a gap to the degenerate 
triplets given by $\Delta=J$. For $N$  odd, the ground
state is described by degenerate spin $\frac 1 2$ 
states with a gap to the quadruplet spin $\frac 3 2$ given
by $\Delta= 1.5 J$.  
In order to mimic  the quasi degenerate nature of the low  energy sector of 
the Kagome spin system, we add a small
disorder term to the magnetic interactions $J_{ij} = J(1+ \delta_{ij})$
where $\delta_{ij} << 1$ are  normally distributed random  numbers.  
This small change in couplings is sufficient to lift the
exact degeneracy of the states, but not
large enough to destroy the energy separation between the low-lying
sectors.
The number of quasi degenerate states of the ground-energy sector is  
macroscopic and given by $A_{even}\approx 1.68^N$ for $N$ even, 
and $A_{odd}\approx 1.74^N$ for $N$ odd.
This resembles
the scaling seen in the Kagome lattice 
where $A_{even} \approx 1.14^N$ for $N$ even and $A_{odd} \approx 1.16^N$ 
for $N$ odd.
Thus, in the $N$ even (odd) case both models feature a quasi-degenerate 
and macroscopically large lowest energy sector  with $S=0$ ($S=1/2$),
and a first excited sector of similar characteristics
with $S=1$ ($S=3/2$). 
We solve the model using exact diagonalization methods \cite{lili-prl,lili-prb} and calculate
the static spin susceptibility and the local dynamic susceptibility in the presence and
absence of an external magnetic field. These calculations are done at finite temperatures
which restricts the size of the systems that we can access numerically.
Nonetheless, as we will show below,
finite size effects are rather minimal. This is partly due to the 
mean field nature of the model, and, also to the fact that
the computed observables are obtained as direct averages over
realizations of disorder \cite{lili-prl,lili-prb}.

Regarding the current debate  on whether 
 the observed experimental behavior is intrinsic to the putative spin liquid state
of a Kagome lattice or due to impurity effects,
 we shall show  below that many experimental observations
can be naturally explained as due to strongly 
interacting many-body spin-$\frac 1 2$ states 
that are present only in $N$ odd systems. We shall argue that
this  leads to the interpretation that   low
energy features seen in current experiments on the paracatamite are controlled by impurities.

\begin{figure}
\centerline{\includegraphics[width=0.4\textwidth,angle=0]{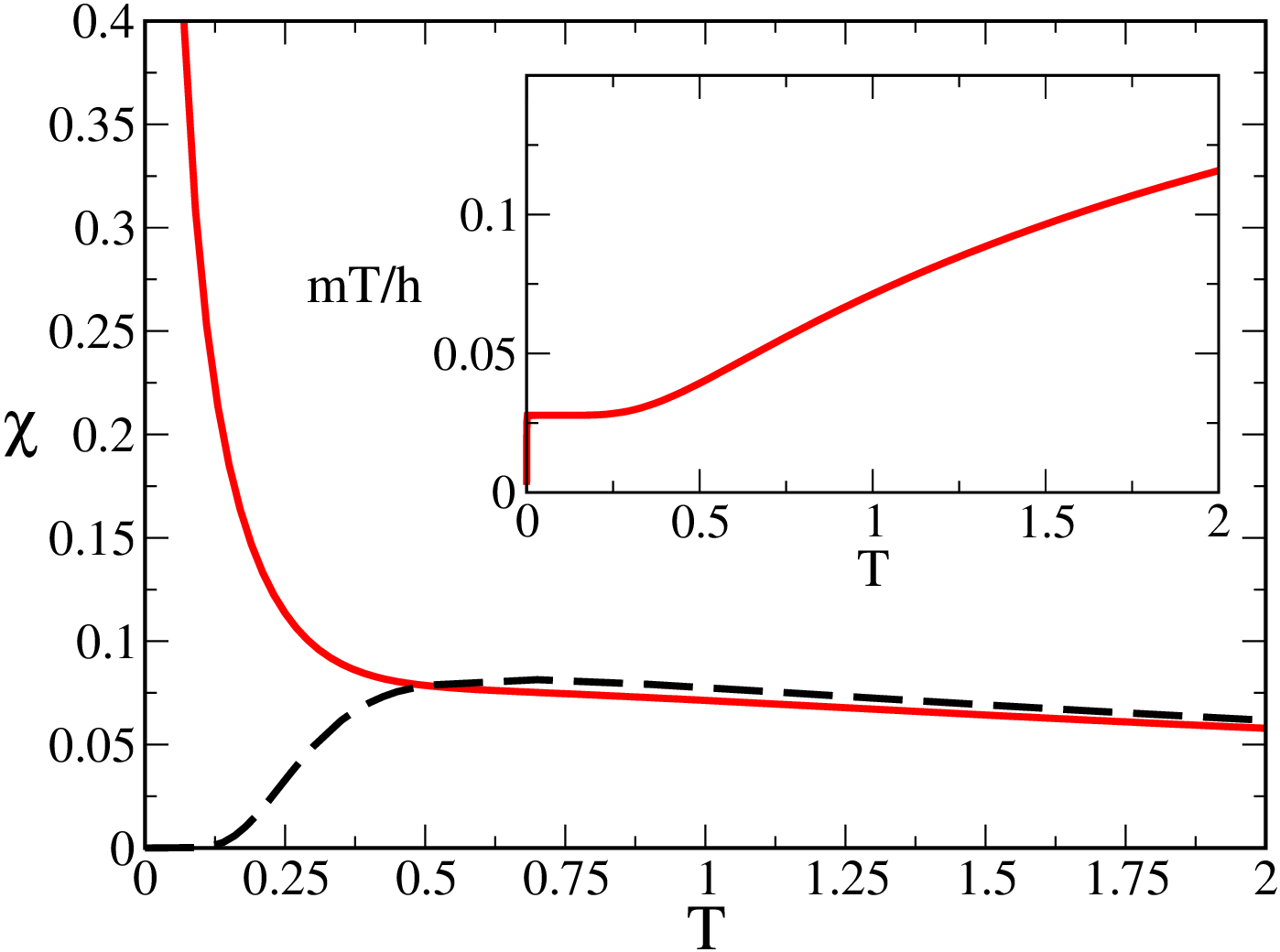}}
\caption{(Color Online) Typical behavior of the static susceptibility as a function of
temperature $T$ for  $N$ odd (full line) and $N$ even (dashed line). The inset shows the typical temperature  variation of the normalized
magnetization in a field for $N$ odd. The finite intercept and the convex shape
of the curve are in qualitative agreement with the experiments of 
Ref.\onlinecite{ofer}
}
\label{fig1}
\end{figure}

\begin{figure}
\centerline{\includegraphics[width=0.4\textwidth,angle=0]{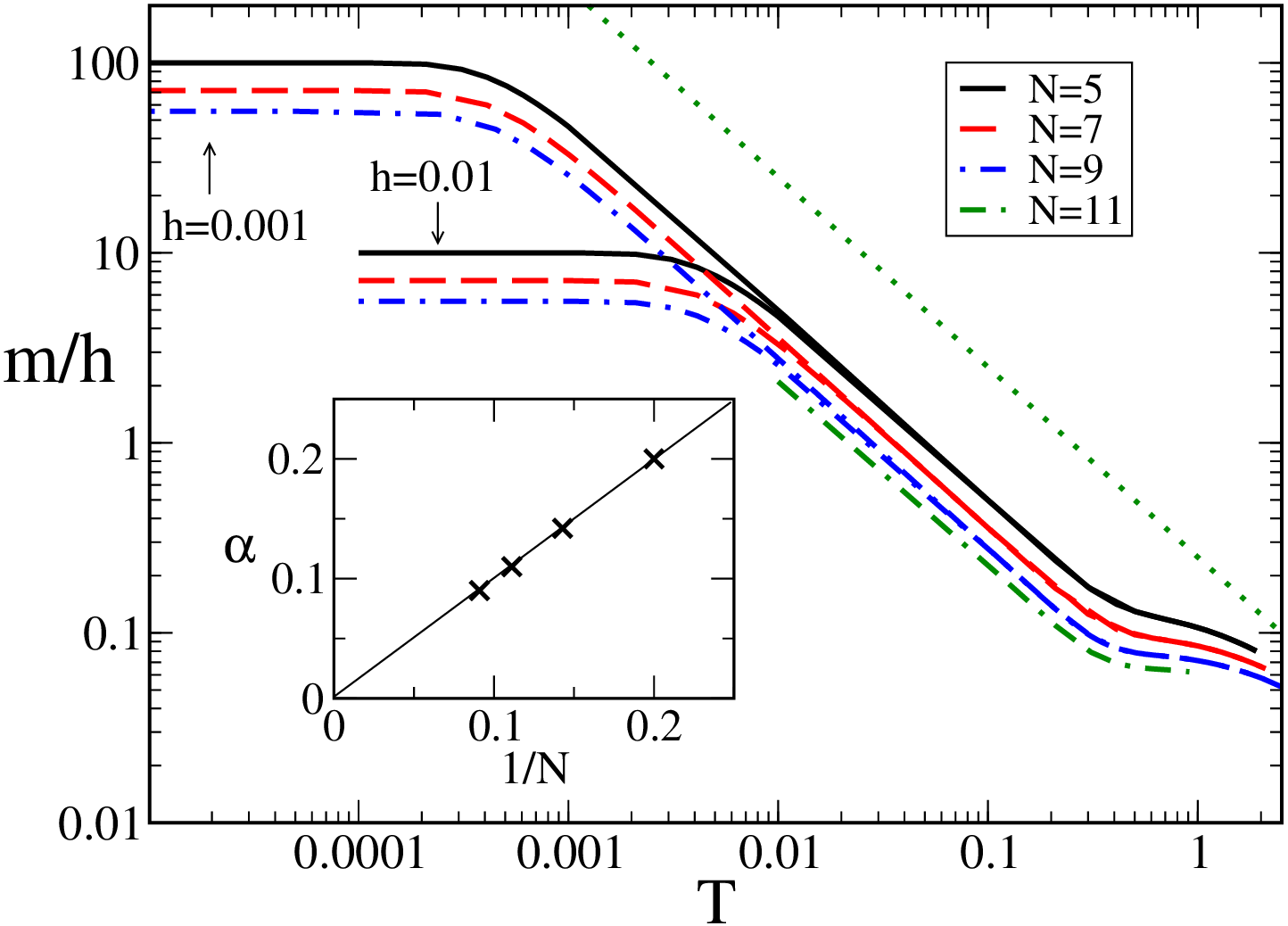}}
\caption{ (Color Online) The finite size scaling for the susceptibility show the
saturation of $\chi$ as $T \to 0$ in a field.  The dotted green line represents
$\chi =1/4T$. The inset shows the variation
of the coefficient $\alpha$ defined in Eq.(2) with system size.
}
\label{fig2}
\end{figure}

To begin with the discussion of our toy model results, we first consider  
the  uniform  static susceptibility per spin, $\chi(T)$. 
This quantity is easily computed from 
the magnetic moment $m(h)$  for small fields i.e., 
$\chi$=$ m/h $, where $m$ is the magnetic 
moment per spin at a temperature $T$ and $h$ is a small
external applied field.   Due to the mean-field nature  and the
translationally invariant spin liquid ground state of our model, the local and the global
magnetic moment are the same. Consequently, the ensuing analysis holds
for local as well as global susceptibilities \cite{ofer}.
From Fig. \ref{fig1}, we see that
$\chi(T)$ is strongly dependent on
whether the total number of spins is  even or odd. This can be easily
understood from the fact that even systems have a singlet ground state
(with other singlets nearly degenerate with the ground state) and a
well defined gap  to the first triplet. This gap leads to
$\chi(T) \propto exp(-\Delta/T)\to 0$ as $T \to 0$. On the other hand, systems with
$N$ odd have a multitude of low-lying spin doublet states that are
quasi-degenerate with the groundstate. Thus, 
as $T \to 0$, this results in a Curie like behavior of the
susceptibility
 \begin{equation}
\chi(T) \approx \frac{\alpha(N)}{4T}
\end{equation}
The inset of Fig.\ref{fig2} shows  the scaling of the Curie coefficient $\alpha(N)$.
Interestingly, the finite size scaling reveals that, despite the
macroscopic number of low-lying spin doublet states, 
$\alpha(N)=1/N$. Thus, the strength of the divergent
part of $\chi(T)$ corresponds exactly to the contribution of a
single effective spin-$\frac 1 2 $, independently of $N$. 
In other words, since $\chi(T)$ is the susceptibility
per spin, our result implies that for systems of any size $N$ (odd) the
strength of the Curie-Weiss contribution of the
{\em total} spin susceptibility $N \chi(T)$ is exactly 1.
This is a clear indication   of a spin-$\frac 1 2$ many-body state
emerging in odd-$N$ systems, which can be also thought of as
a missing spin-$\frac 1 2$ (defect) in an otherwise $S=0$ spin liquid.
The existence of this effective ''dangling'' spin is further 
clarified by the  behavior of $\chi(T)$ at small finite fields $h$. 
From  Fig.\ref{fig2}, we see that the Curie divergency
observed in odd $N$  systems is cut-off by $h$, and $\chi$
becomes temperature independent for $T \lesssim h$. This is in agreement with the results reported in Ref.\onlinecite{ofer}, where
a field of 0.2T was applied and the saturation of $\chi(T)$ below 0.2K
was observed.
 At intermediate and high temperatures, $\chi(T)$ is almost
the same for systems with odd/even $N$ modulo finite size scaling
effects.
At high enough temperatures $ T > J$, we recover the usual 
Curie-Weiss
behavior of the susceptibility $ \chi(T) \propto (T+ T_{\rm CW})^{-1}$. As
expected for a mean field model where every spin has $N$-1 neighbors, 
we find  that the Curie-Weiss
temperature which fixes the scale of the antiferromagnetic exchange in
the system is given by $T_{\rm CW}\approx$-0.3(N-1)$J$ where the factor 0.3
is due to quantum fluctuations and the inherent frustration. To compare the scale
of the Curie-Weiss temperature  predicted by the mean field model 
with that of real experiments, we can
use the experimental value of $J$=170K. For Kagome lattices, since
the number of nearest neighbors is $4$, we obtain 
$\vert T_{\rm CW}Ê\vert$ =$1.2$J ($\simeq$204K) which is of  the same order as 
the experimental estimate of around
270$\sim$320K. 

\begin{figure} 
\centerline{\includegraphics[width=0.4\textwidth,angle=0]{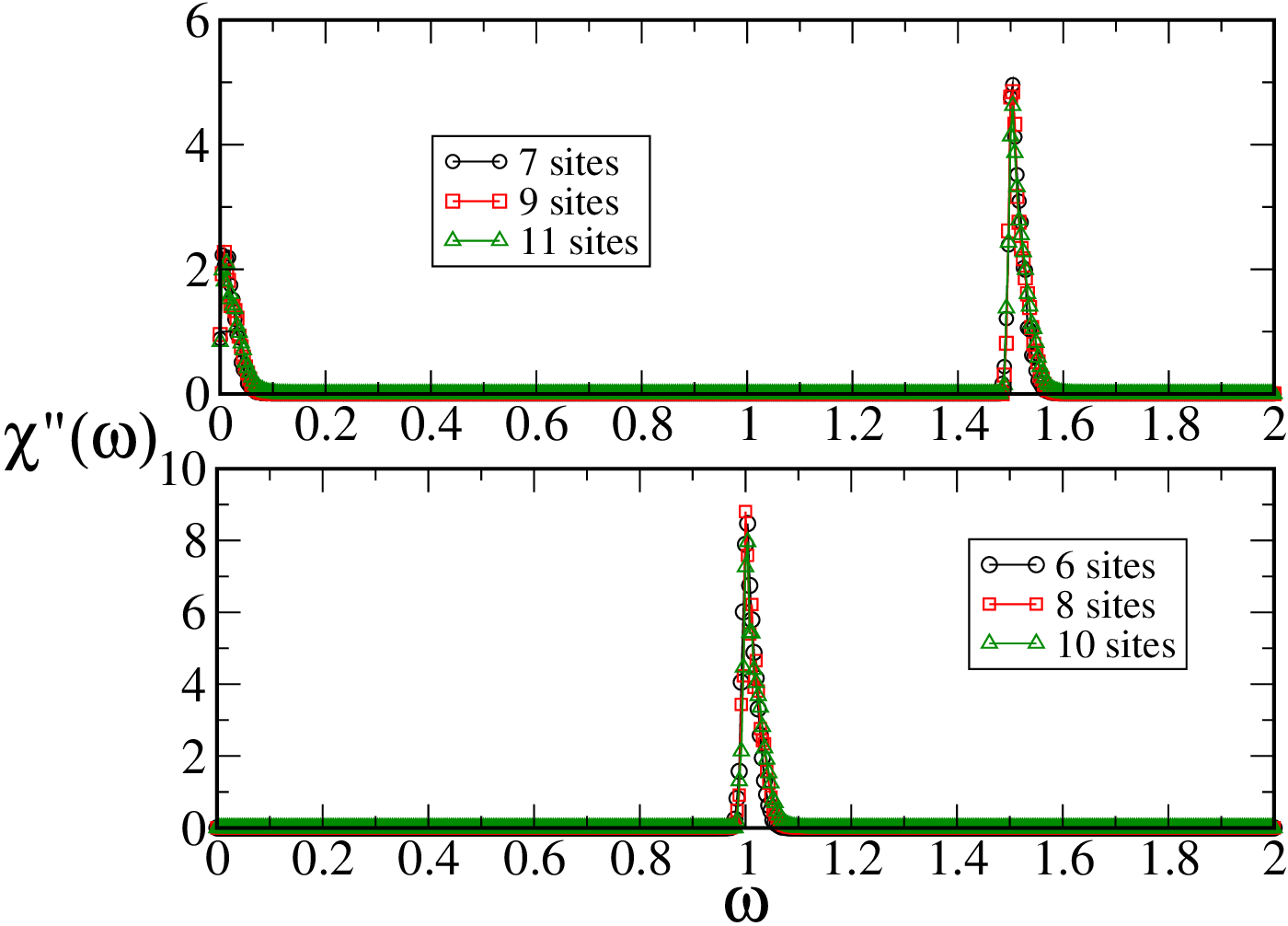}}
\caption{(Color Online) The $T$=0 dynamic susceptibility $\chi''(\omega)$ as a function of
frequency $\omega$ in units of $J$ for systems with $N$ odd and $N$ even.
}
\label{fig3}
\end{figure}

We now  discuss the behavior of dynamical quantities.
We use the  following spectral decomposition to calculate the
imaginary part of the dynamic susceptibility $Im\chi\equiv\chi''$:
\begin{widetext}
\begin{equation}
\chi''(\omega)= 
\frac {\pi}{ZNM}  
\sum_{m=1}^M \sum_{i=1}^N \sum_{\nu\mu}\vert\langle \mu^{m}\vert S_i^z\vert \nu^{m} 
\rangle \vert^2 \delta(\omega- E_{\mu}^{m}+E_{\nu}^{m}) \exp(-\beta E_{\nu}^{m}) 
[1 - \exp(-\beta \omega)]
\label{chi}
\end{equation}
\end{widetext}
where $Z$ is the partition function  
and $M$ the number of disorder realizations that  are used to average $\chi$.
In Fig. \ref{fig3} we
plot $\chi''$ for both even and odd systems.
Note that, as  already anticipated, 
the finite size effects on $\chi''(\omega)$ are quite weak. 
Both even and odd spectra show the presence of a prominent peak
at a frequency of order of the gap $\Delta \propto J$ to the first excited state.  Note that in the absence of disorder, these peaks are Delta functions.
The peak at $\omega_0 \approx J$ for even systems results form excitations between states in the
$S_{tot}$=0 and  $S_{tot}$=1 sectors, while in the case of odd systems the transitions are 
between $S_{tot}$=1/2 and  $S_{tot}$=3/2 with an $\omega_0 \approx 3J/2$ \cite{note}. 
However, we note that there is a   fundamental  difference between     $N$ even and odd  
in that  the latter   has   significant low frequency spectral weight. 
In the rest of the paper,  we shall discuss   the   behavior of this low frequency  feature
in the context of   recent neutron scattering
experiments on ZnCu$_{3}$(OH)$_6$Cl$_2$ \cite{helton}.

\begin{figure}
\centerline{\includegraphics[width=0.4\textwidth,angle=0]{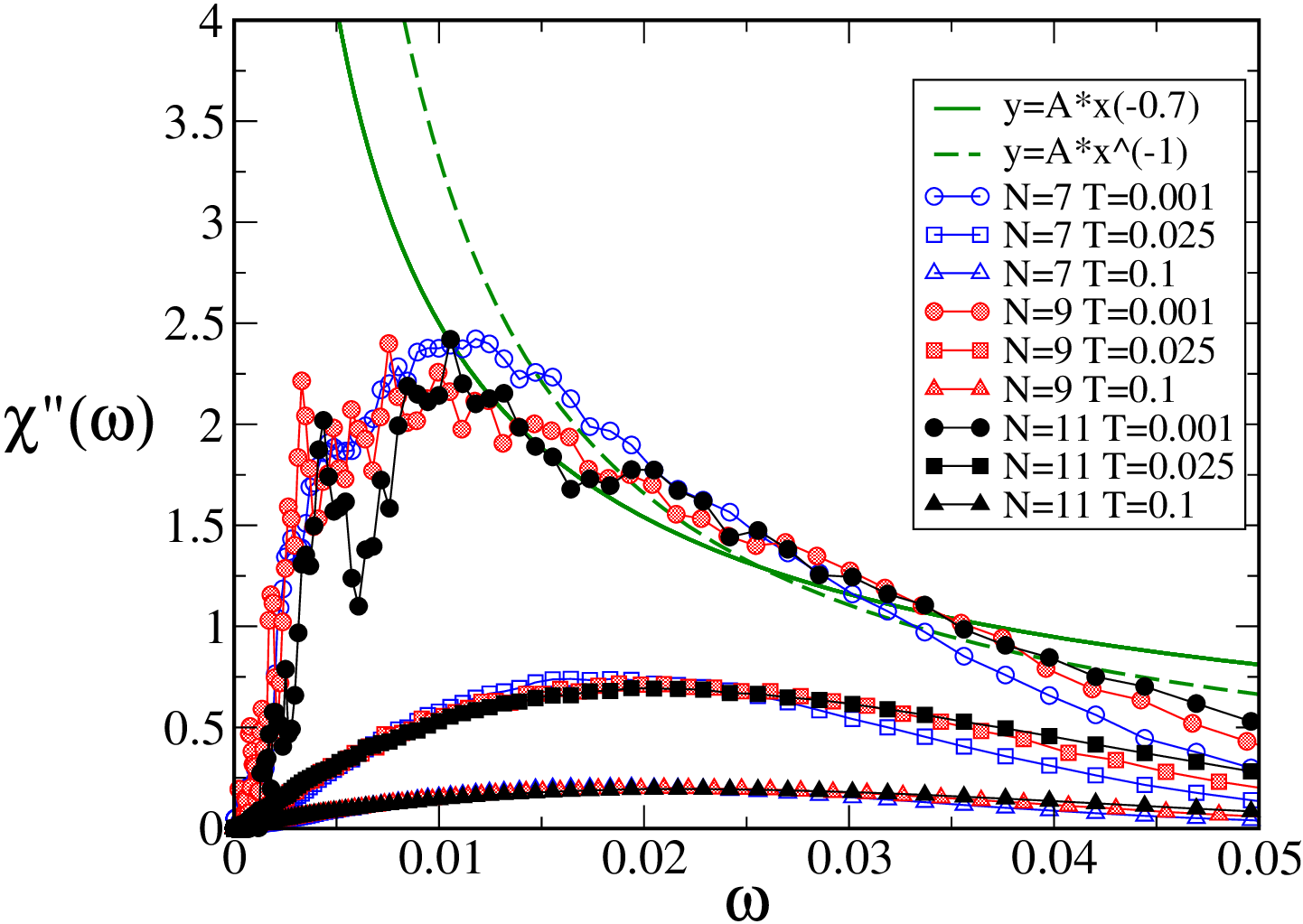}}
\caption{ (Color Online) The low frequency peak in $\chi''(\omega)$ as a function of
frequency $\omega$ in units of $J$ for different temperatures $T$=
0.001, 0.025 and 0.1$J$. 
Taking $J$=170K, our data approximately correspond to the
same experimental window reported in Ref.\onlinecite{helton}.
The solid and dashed lines represent the power law fits proposed
in that experimental study.} 
\label{fig4}
\end{figure}

The qualitative difference between the low
energy spectra of even and odd systems can be readily understood.
This spectra 
corresponds to transitions within the lowest energy sector that
contains a macroscopic number of quasi-degenerate states (i.e., within
energy {\it O}($\delta J$) from the groundstate). This sector has
$S_{tot}$=0 for even systems and $S_{tot}$=1/2 for odd systems.
Since the spin susceptibility induces transitions between states
which differ in one unit of  the $z$ component of the spin, these are forbidden
 between  the singlets of even size systems, but are perfectly
possible between the low-lying doublets of  odd systems.
Based on this argument, the low frequency feature 
in $\chi''(\omega)$ seen in odd systems should have a characteristic
width of order $\delta J$ as  is verified by Figs.\ref{fig3} and \ref{fig4}.
Thus, the physics of the low frequency contribution to the dynamical
susceptibility is also due to the same effective  spin-$\frac  1 2$ degree of freedom
that was previously responsible for the low temperature contribution to
$\chi(T)$.

We now focus on  the temperature dependence of this low frequency  peak.
As one increases $T$ from $10^{-3}J$ to $10^{-1}J$, which is roughly the
same range as  that studied in the experiments of  Ref.\onlinecite{helton},
we observe the following:  first, the height and
intensity of the peak diminishes  dramatically and secondly,  the peak shifts  to higher frequencies.
These behaviors are qualitatively similar to those
seen in neutron experiments \cite {helton}.
At smallest $T$ there is a strong enhancement of $\chi''(\omega)$ 
towards lower frequencies. However, eventually the susceptibility
vanishes as $\omega \to 0$ at the very low frequency end of
the spectra. The curves for the three different system sizes shown in the figure
demonstrate that even in this low frequency regime, the finite size
effects are rather  small.
As argued before, the width of the peak is controlled by the 
disorder distribution $\delta$ which is a phenomenological
parameter in our model. Since $J$=170K, we adjust 
the disorder strength to the value $\delta=0.01$ to fit the 
experimental neutron data. 
The frequency range shown in Fig.\ref{fig4} corresponds to the
experimental range of the reported neutron scattering \cite{helton}
and we clearly see that taking into account the experimental error bars, 
our results are indeed compatible with the observed data. 
For comparison, the power law fits suggested by the experiments
are also shown in this figure.  However, our results do not predict
a divergence of  $\chi''(\omega)$ for $\omega \to 0$. Therefore,
our study suggest that the putative divergence observed  
may be merely due to rather large error bars
of the lowest frequency data, which is 
the region where the experiment suffers from largest uncertainty.
Moreover,  the lack of finite size effects  permits us  to obtain a rather reliable estimate of the integrated intensity
of the low frequency feature, and we find that it amounts to about
25\% of the total spectral weight. This value is compatible
with the reported 20\% estimated from the neutron experiments.

\begin{figure}
\centerline{\includegraphics[width=0.4\textwidth,angle=0]{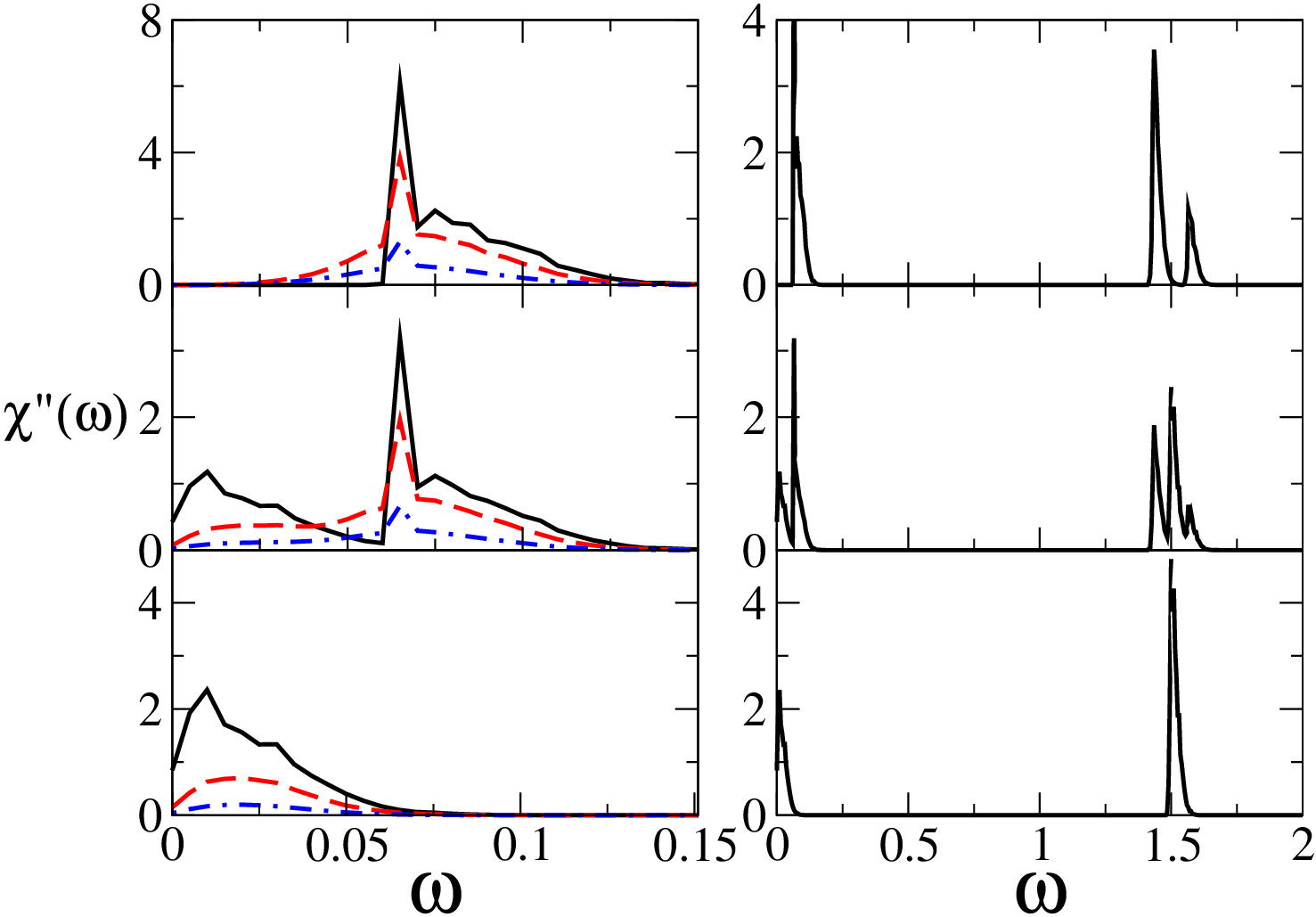}}
\caption{ (Color Online) $\chi''(\omega)$ as a function of
frequency $\omega$. An external magnetic field is applied along
the transverse $x$-axis (top), the longitudinal $z$-axis (bottom) and
a tilted axis in the $x-z$ plane (middle).
To match the experimental value we set $h$=0.067$J$, and $T$=0.001, 
0.025 and 0.1$J$ (top, middle and bottom curve in each left panel).
The left column panels show a detailed of the low energy region and
the corresponding full spectra are shown in the right column (only the
lowest $T$ is plotted).
}
\label{fig5}
\end{figure}

We finally consider the effect of a magnetic field on
the dynamical susceptibility. 
In the presence of an external field ${\bf h}= h {\bf z}$, $\chi$ 
is the longitudinal susceptibility, whereas
for fields aligned along the $x-$direction, $\chi$ represents the 
transverse susceptibility.
In   we plot  the
temperature evolution of $\chi''(\omega)$
for  applied fields
with different orientations.
The strength of the magnetic field is set to $h=0.067 J$ so as to 
match the experimental value of $11.5$Tesla used in Ref.\onlinecite{helton}. 
Our results show that $h$ has a strong effect on 
$\chi''(\omega)$ at both high and low frequencies.
For an applied field in the $z$-direction, the longitudinal susceptibility $\chi''$ (bottom panels in Fig.\ref{fig5}),
 remains essentially unchanged
since a longitudinal field does not produce spin flips.
For a transverse field, the computed $\chi''(\omega)$  now corresponds
 to the transverse susceptibility and as expected, the high frequency 
feature at $\omega_0=3/2J$ shows the usual Zeeman splitting  
$\omega_0 \pm h$ (top right panel).
This splitting corresponds to excitations
from states with $S^z_{tot}= \pm 1/2$ to states with 
$ S^z_{tot}=\pm 3/2$. Interestingly,  we also see a shift of the low energy 
part of the spectrum towards a higher frequency $\omega \sim h$.
A single spin-$\frac 12$ would give a sharp peak at the
Larmor frequency $h$. However, in the present case, in
addition to a sharp peak, we
also find a broad asymmetric contribution. This anomalous behavior is due to the
quasi-degeneracy of the low-lying states and the many-body nature of the 
remnant spin-$\frac 12$ degree of freedom present in odd systems.
Interestingly, a similar broad anomalous feature 
was also seen in the reported neutron data \cite{helton}. 
For fields applied in the tilted (1,0,1) direction (central panels) the resulting 
response can be easily interpreted as the superposition of
the transverse and longitudinal responses. Thus, the high energy feature is split into three peaks centered
around $\omega_0$ and $\omega_0\pm h$, and a similar analysis
is also valid for low frequencies. The response
in a tilted field  can be considered as a generic case, and is also more likely to 
be closer to the experiments which were performed on powder samples.
 
We believe that the accord we find between
our mean field toy model results and experiments may provide
useful guidance,
especially, in regard to the  current debate on whether the low energy behavior seen in
experiments is 
dominated by impurity effects.   
In fact, odd systems can be viewed as even 
systems with a spin defect (i.e., an impurity), or, alternatively,
as a single (impurity) spin dressed by the coupling to 
an $S_{tot}$=0 (even) spin liquid. 
Thus,  our results for odd systems naturally lead to the 
interpretation of the anomalies observed in
current experiments as arising
from impurities/defects which
may  originate from the 
lack of precise stoichiometric control in currently
available samples.  
Once a chemical handle is found for the synthesis of the compounds, 
a systematic study of the dependence with
Zn-Cu substitution may help disentangle the impurity and  intrinsic
contributions, and eventually validate our proposed scenario for
this fascinating quantum magnet system.

We thank C. Lhuillier for introducing  us to this problem.

\date{\today}

\begin{thebibliography}{10}
\bibitem{note} The fact that these excitation energies 
do not converge with even and odd $N$ as $N \to \infty$ is
a peculiarity of the model which is irrelevant for the
discussion of the present work.
\end{thebibliography}
\end{document}